\documentclass[aps,preprint]{revtex4}
\usepackage{amssymb,epsf}
\usepackage{amsmath}
\usepackage{graphicx}

\begin{document}

\title{Magnetic Branes in Third Order Lovelock-Born-Infeld Gravity}
\author{M. H. Dehghani$^{1,2}$\footnote{email address:
mhd@shirazu.ac.ir} N. Bostani$^{1}$ and S. H.
Hendi$^{3}$\footnote{hendi@mail.yu.ac.ir}}

\affiliation{$^1$ Physics Department and Biruni Observatory, College of Sciences, Shiraz University, Shiraz 71454, Iran\\
         $^2$ Research Institute for Astrophysics and Astronomy of Maragha (RIAAM), Maragha,
         Iran\\
         $^3$ Physics Department, College of Sciences, Yasouj University, Yasouj 75914, Iran}

\begin{abstract}
Considering both the nonlinear invariant terms constructed by
the electromagnetic field and the Riemann tensor in  gravity action, we
obtain a new class of $(n+1)$-dimensional magnetic brane solutions
in third order Lovelock-Born-Infeld gravity. This class of
solutions yields a spacetime with a longitudinal nonlinear
magnetic field generated by a static source. These solutions have
no curvature singularity and no horizons but have a conic geometry
with a deficit angle $\delta$. We find that, as the Born-Infeld
parameter decreases, which is a measure of the increase of the
nonlinearity of the electromagnetic field, the deficit angle
increases. We generalize this class of solutions to the case of
spinning magnetic solutions and find that, when one or more
rotation parameters are nonzero, the brane has a net electric
charge which is proportional to the magnitude of the rotation
parameters. Finally, we use the counterterm method in third order
Lovelock gravity and compute the conserved quantities of these
spacetimes. We found that the conserved quantities do not depend
on the Born-Infeld parameter, which is evident from the fact that
the effects of the nonlinearity of the electromagnetic fields on
the boundary at infinity are wiped away. We also find that the
properties of our solution, such as deficit angle, are independent
of Lovelock coefficients.
\end{abstract}

\maketitle
\section{Introduction}

Actions of Lovelock gravity and Born-Infeld electrodynamics have been the
subject of wide interest in recent years. This is due to the fact that both
of them emerge in the low-energy limit of string theory \cite{frad}%
. On the gravity side, string theories in their low-energy limit give rise
to effective models of gravity in higher dimensions which involve higher
curvature terms, while on the electrodynamics side the effective action for
the open string ending on D-branes can be written in a Born-Infeld form.
While Lovelock gravity \cite{Lov} was proposed to have field equations
with at most second order derivatives of the metric \cite{Zum}, the
nonlinear electrodynamics was proposed, by Born and Infeld, with the aim of
obtaining a finite value for the self-energy of a pointlike charge \cite
{BI}. Lovelock gravity reduces to Einstein gravity in four dimensions
and also in the weak field limit, while the Lagrangian of the Born-Infeld
(BI) electrodynamics reduces to the Maxwell Lagrangian in the weak field
limit. There have been considerable works on both of these theories. In
Lovelock gravity, there have been some attempts for understanding the role of
the higher curvature terms from various points of view. For example, exact
static spherically symmetric black hole solutions of the second order
Lovelock gravity have been found in Ref. \cite{Des}, and of the
Gauss-Bonnet-Maxwell model in Ref. \cite{Wil1}. An exact static solution in
third order Lovelock gravity was introduced in Ref. \cite{Deh2} and of the charged
rotating black brane solution in Ref. \cite{DM1}, and the magnetic solutions
with longitudinal and angular magnetic field were considered in Ref. \cite{DB}.
All of these works were in the presence of a linear electromagnetic field. The
first attempt to relate the nonlinear electrodynamics and gravity has been done
by Hoffmann \cite{Hoffmann}. He obtained a solution of the Einstein
equations for a pointlike Born-Infeld charge, which is devoid of the
divergence of the metric at the origin that characterizes the
Reissner-Nordstr\"{o}m solution. However, a conical singularity remained
there, as it was later objected by Einstein and Rosen. The spherically
symmetric solutions in Einstein-Born-Infeld gravity with or without a
cosmological constant have been considered by many authors \cite{EBI,Dey}.

In this paper we are dealing with the issue of the spacetimes
generated by static and spinning brane sources which are
horizonless and have non-trivial external solutions. These kinds
of solutions have been investigated by many authors in four
dimensions. Static uncharged cylindrically symmetric solutions of
Einstein gravity in four dimensions were considered in Ref. \cite
{Levi}. Similar static solutions in the context of cosmic string
theory were found in Ref. \cite{Vil}. All of these solutions
\cite{Levi,Vil} are horizonless and have a conical geometry; they
are everywhere flat except at the location of the line source. The
extension to include the electromagnetic field has also been done
\cite{Muk,Lem1}. In three dimensions, the rotating magnetic
solutions of Einstein gravity have been studied in Ref. \cite{Lem2},
while those of Brans-Dicke gravity have been considered in
Ref. \cite{Lem3}. The generalization of these solutions in Einstein
gravity in the presence of a dilaton and Born-Infeld electromagnetic
field has been done in Ref. \cite{DSH}. Our aim in this paper is to
construct $(n+1)$-dimensional horizonless solutions of third order
Lovelock gravity in the presence of a nonlinear electromagnetic
field and investigate the effects of the nonlinearity of the
electromagnetic field on the properties of the spacetime such as
the deficit angle of the spacetime. The first reason for studying
higher-dimensional solutions of gravity, which invite higher order
Lovelock terms, is due to the scenario of string theory and brane
cosmology. The idea of brane cosmology, which is also consistent
with string theory, suggests that matter and gauge interaction
(described by an open string) may be localized on a brane embedded
into a higher-dimensional spacetime, and all gravitational objects
are higher-dimensional \cite{Ran, Dval}. The second reason for
studying higher-dimensional solutions is the fact that
four-dimensional solutions have a number of remarkable properties.
It is natural to ask whether these properties are general features
of the solutions or whether they crucially depend on the world
being four-dimensional. Finally, we want to check whether the
counterterm method introduced in Refs. \cite{DBSH, DM1} can be applied
to the case of solutions in the presence of a nonlinear
electromagnetic field, too.

The outline of our paper is as follows. We give a brief review of
the field equations of Lovelock gravity in the presence of
a Born-Infeld electromagnetic field in Sec. \ref{Fiel}. In Sec.
\ref{Long} we present static horizonless solutions which produce
a longitudinal magnetic field and investigate the effects of
the nonlinearity of the electromagnetic field and Lovelock terms on
the deficit angle of the spacetime. Section \ref{Rot} will be
devoted to the generalization of these solutions to the case of
rotating solutions and use of the counterterm method to compute
the conserved quantities of them. We finish our paper with some
concluding remarks.

\section{Field equations\label{Fiel}}

A natural generalization of general relativity in higher-dimensional
spacetimes with the assumption of Einstein, that the
left-hand side of the field equations is the most general
symmetric conserved tensor containing no more than second
derivatives of the metric, is Lovelock theory. Lovelock found the
most general symmetric conserved tensor satisfying this property.
The resultant tensor is nonlinear in the Riemann tensor and
differs from the Einstein tensor only if the spacetime has more
than 4 dimensions. The Lovelock tensor in $(n+1)$ dimensions may
be written as \cite{Lov}
\begin{equation}
\sum_{i=1}^{[n/2]}\alpha'_{i}[H_{\mu \nu }^{(i)}-\frac{1}{2}g_{\mu \nu }%
{\cal L}^{(i)}]+\Lambda g_{\mu \nu }=8\pi T_{\mu \nu },  \label{FF}
\end{equation}
where $\Lambda =-n(n-1)/2l^{2}$ is the cosmological constant for
asymptotically anti-de Sitter (AdS) solutions, $[n/2]$ denotes the integer part of $n/2$,
$\alpha'_1=1$, $\alpha'_{i}$'s for $i\geq 2$ are Lovelock coefficients, $T_{\mu \nu }$ is the
energy-momentum tensor, and ${\cal H}_{\mu \nu }^{(i)}$ and ${\cal L}^{(i)}$
are
\begin{eqnarray}
&&{\cal H}_{\mu \nu }^{(i)}=\frac{i}{2^{i}}\delta _{\mu \text{ }\nu_{2}
\cdots \nu _{2i}}^{\mu _{1}\mu _{2}\cdots \mu _{2i}}R_{\mu _{1}\mu
_{2}\nu }^{\phantom{\alpha_1\alpha_2\nu}{\nu_2}}R_{\mu _{3}\mu _{4}}^{%
\phantom{\alpha_3\alpha_4}{\nu_3\nu_4}}\cdots R_{\mu _{2i-1}\mu _{2i}}^{%
\phantom{\alpha_2i-1 \alpha_{2i}}{\nu_{2i-1} \nu_{2i}}},  \label{FF1} \\
&&{\cal L}^{(i)}=\frac{1}{2^{i}}\delta _{\nu_1\text{ }\nu _{2}\cdots \nu
_{2i}}^{\mu _{1}\mu _{2}\cdots \mu _{2i}}R_{\mu _{1}\mu _{2}}^{%
\phantom{\alpha_1\alpha_2}{\nu_1\nu_2}}\cdots R_{\mu _{2i-1}\mu _{2i}}^{%
\phantom{\alpha_{2i-1} \alpha_{2i}}{\nu_{2i-1} \nu_{2i}}}.  \label{FF2}
\end{eqnarray}

In this paper, we consider up to third order Lovelock terms. The
first term is just the Einstein tensor, and the second and third
terms are Gauss-Bonnet and third order Lovelock tensors, respectively, which are
written in terms of Riemann tensor explicitly in Ref. \cite{Hoi}. In
the presence of a Born-Infeld electromagnetic field, the
energy-momentum tensor of Eq. (\ref{FF}) is
\begin{equation}
T_{\mu \nu }=\frac{1}{4\pi }\left(\frac{1}{4}g_{\mu \nu
}L(F)+\frac{F_{\mu \lambda }F_{{\nu }}\,^{\lambda }}{\sqrt{1+\frac{F^{2}}{%
2\beta ^{2}}}}\right),  \label{TT}
\end{equation}
where $F_{\mu \nu }$ is the nonlinear electromagnetic field tensor which
satisfies the BI equation
\begin{equation}
\partial _{\mu }\left( \frac{\sqrt{-g}F^{\mu \nu }}{\sqrt{1+\frac{F^{2}}{%
2\beta ^{2}}}}\right) =0,  \label{BI}
\end{equation}
and $F^{2}=F^{\mu \nu }F_{\mu \nu }$. The parameter $\beta $ is called
the Born-Infeld parameter which has the dimension of $ length^{-2}$, and, as it goes to $%
\infty $, the BI equation reduces to the standard Maxwell equation.

\section{Static magnetic branes\label{Long}}
Here we want to obtain the $(n+1)$-dimensional solutions of Eqs. (\ref{FF}-%
\ref{BI}) which produce longitudinal magnetic fields in the Euclidean
submanifold spans by $x^{i}$\ coordinates ($i=1,...,n-2$). We will work with
the following ansatz for the metric \cite{Lem1}:
\begin{equation}
ds^{2}=-\frac{\rho ^{2}}{l^{2}}dt^{2}+\frac{d\rho ^{2}}{f(\rho )}%
+l^{2}f(\rho )d\phi ^{2}+\frac{\rho ^{2}}{l^{2}}dX^{2},  \label{Met1a}
\end{equation}
where $dX^{2}={{\sum_{i=1}^{n-2}}}(dx^{i})^{2}$ is the Euclidean metric on
the $(n-2)$-dimensional submanifold. The angular coordinate $\phi $ is
dimensionless as usual and ranges in $[0,2\pi ]$, while $x^{i}$'s range in $%
(-\infty ,\infty )$. The motivation for this metric gauge $[g_{tt}\varpropto
-\rho ^{2}$ and $(g_{\rho \rho })^{-1}\varpropto g_{\phi \phi }]$ instead of
the usual Schwarzschild gauge $[(g_{\rho \rho })^{-1}\varpropto g_{tt}$ and $%
g_{\phi \phi }\varpropto \rho ^{2}]$ comes from the fact that we are looking
for a horizonless solution.

The electromagnetic field equation (\ref{BI}) can be integrated immediately
to give
\begin{equation}
F_{\phi \rho }=\frac{2ql^{n-1}}{\rho ^{n-1}\sqrt{1-\eta }},  \label{Fpr}
\end{equation}
where $q$ is the charge parameter of the solution and
\begin{equation}
\eta =\frac{4q^{2}l^{2n-4}}{\beta ^{2}\rho ^{2(n-1)}}.  \label{eta}
\end{equation}
Equation (\ref{Fpr}) shows that $\rho $\ should be greater than $\rho
_{0}=(2ql^{n-2}\beta )^{1/(n-1)}$\ in order to have a real nonlinear
electromagnetic field and consequently a real spacetime. To find the
function $f(\rho )$, one may use any components of Eq. (\ref{FF}). The
simplest equation is the $\rho \rho $ component of these equations which can
be written as
\begin{eqnarray}
&&\left( \alpha_{3}\rho f^{2} -2 \alpha_{2}\rho ^{3} f+\rho ^{5}\right)
f^{\prime } +\frac{n-6}{3}\alpha_{3}f^{3}-(n-4)\alpha_{2}\rho ^{2}f^{2},  \nonumber \\
&&+(n-2)\rho ^{4}f-
\frac{n}{l^{2}}\rho ^{6}=\frac{%
\beta ^{2}\rho ^{6}}{2(n-1)}(1-\sqrt{1-\eta }),  \label{Eqf}
\end{eqnarray}
where the prime denotes the derivative with respect to $\rho $ and
\begin{equation}
\alpha _{2}=(n-2)(n-3) \alpha'_{2},\quad \alpha _{3}=72(_{\phantom{n}{4}}^{n-2}) \alpha'_{3}.
\end{equation}
The only real solution of Eq. (\ref{Eqf}) is
\begin{equation}
f(\rho )=\frac{\alpha_2 \rho ^{2}}{\alpha_{3}}\left\{1-
\left(\sqrt{\gamma +j^{2}(\rho )}+j(\rho )\right) ^{1/3}+\gamma
^{1/3}\left( \sqrt{\gamma +j^{2}(\rho )}+j(\rho )\right)
^{-1/3}\right\} ,  \label{f1}
\end{equation}
where
\begin{eqnarray}
\gamma &=&(\frac{\alpha_{3}}{\alpha_{2}^{2}}-1)^{3},\\
j(\rho) &=&-1+\frac{3\alpha_3}{2\alpha^2_2} -\frac{3\alpha_{3}^{2}}{2l^2\alpha_2^3}k(\rho), \label{jrho} \\
k(\rho)&=&1+\frac{2ml^3}{\rho^{n}}+%
\frac{\beta ^{2}l^2 h(\eta )}{2 n(n-1)} .  \label{krho}
\end{eqnarray}
The constants $m$ and $q$ in Eq. (\ref{krho}) are the mass and charge
parameters of the metric, respectively, which are related to the mass and charge density of
the solution, and $h(\eta )$ is given as
\begin{equation}
h(\eta )=1-\sqrt{1-\eta }-\frac{(n-1)\eta}{(n-2)}\, _{2}F_{1}\left( \left[
\frac{1}{2},\frac{n-2}{2n-2}\right] ,\left[ \frac{3n-4}{2n-2}\right] ,\eta
\right) ,
\end{equation}
where $_{2}F_{1}([a,b],[c],z)$ is hypergeometric function. Using the fact
that $_{2}F_{1}([a,b],[c],z)$ has a convergent series expansion for $|z|<1$%
, we can find the behavior of the metric for large $\rho $\ and $\beta $.
The function $j(\rho )$ approaches the constant
\[ \lambda=-1+\frac{3 \alpha_3}{2 \alpha_2^2}-\frac{3 \alpha_3^2}{2l^2 \alpha_2^3},\]
 as $\rho $ goes to infinity,
and the effective cosmological constant for the spacetime is
\begin{equation}
\Lambda_{{\rm eff}}=-\frac{n(n-1)\alpha_2}{2 \alpha_{3}}\left\{ 1+{\left( \sqrt{%
\gamma +\lambda ^{2}}+\lambda \right) ^{1/3}-}\gamma ^{1/3}\left( \sqrt{%
\gamma +\lambda ^{2}}+\lambda \right) ^{-1/3}\right\}.
\end{equation}
In the rest of the paper, we investigate only the case of $\gamma
\geq 0$, for which $f(\rho )$ is real. The function $f(\rho ) $ is
negative for large values of $\rho $, if $\Lambda_{{\rm eff}}>0 $.
Since $g_{\rho \rho }$ and $g_{\phi \phi }$ are related by $f(\rho
)=g_{\rho \rho }^{-1}=l^{-2}g_{\phi \phi }$, when
$g_{\rho \rho }$ becomes negative (which occurs for large $\rho$),
so does $g_{\phi \phi }$. This leads to an apparent change of
signature of the metric from $(n-1)+$ to $(n-2)+$ as $\rho $ goes
to infinity, which is not allowed. Thus, $\Lambda_{{\rm eff}}$
should be negative which occurs provided the Lovelock coefficients
are assumed to be positive.

In order to study the general structure of the solution given in
Eq. (\ref {f1}), we first look for curvature singularities. It is
easy to show that the Kretschmann scalar $R_{\mu \nu \lambda
\kappa }R^{\mu \nu \lambda \kappa }$ diverges at $\rho =0$, and
therefore one might think that there is a curvature singularity
located at $\rho =0$. However, as we will see below, the spacetime
will never achieve $\rho =0$. The function $f(\rho )$ is negative
for $\rho <r_{+}$ and positive for $\rho >r_{+}$, where $r_{+}$ is
the largest real root of $f(\rho )=0$ which reduces to
\begin{equation}
k(r_+)=1 +\frac{2ml^3}{r_+^{n}}+ \frac{\beta ^{2}l^2 h(\eta_+ )}{2
n(n-1)}=0.
\end{equation}
Again $g_{\rho \rho }$ cannot be negative (which occurs for $\rho
<r_{+}$), because of the change of signature of the
metric from $(n-1)+$ to $(n-2)+$. Thus, one cannot extend the spacetime to $%
\rho <r_{+}$. To get rid of this incorrect extension, we introduce
the new radial coordinate $r$ as
\begin{equation}
r^{2}=\rho ^{2}-r_{+}^{2}\Rightarrow d\rho ^{2}=\frac{r^{2}}{r^{2}+r_{+}^{2}}%
dr^{2} \label{Tr1}.
\end{equation}
With this new coordinate, the metric (\ref{Met1a}) is
\begin{equation}
ds^{2}=-\frac{r^{2}+r_{+}^{2}}{l^{2}}dt^{2}+\frac{r^{2}}{%
(r^{2}+r_{+}^{2})f(r)}dr^{2}+l^{2}f(r)d\phi ^{2}+\frac{r^{2}+r_{+}^{2}}{l^{2}%
}dX^{2},  \label{Metr1b}
\end{equation}
where the coordinates $r$ and $\phi $ assume the value $0\leq r<$ $\infty $
and $0\leq \phi <2\pi$, respectively. The function $f(r)$ is now given as
\begin{equation}
f(r)=\frac{\alpha_2(r^{2}+r_{+}^{2})}{\alpha_{3}}\left\{1-
\left(\sqrt{\gamma
+j^{2}(r)}+j(r)\right) ^{1/3}+\gamma ^{1/3}\left( \sqrt{\gamma +j^{2}(r)}%
+j(r)\right) ^{-1/3}\right\} ,  \label{F2}
\end{equation}
where
\begin{eqnarray}
\eta  &=&\frac{4q^{2}l^{2n-4}}{\beta ^{2}(r^{2}+r_{+}^{2})^{(n-1)}},\label{eta2}\\
j(r)&=&-1+\frac{3\alpha_3}{2\alpha^2_2}-\frac{3\alpha_3^2}{2l^2\alpha_2^3}
k(r) \label{jr2}
 \\
k(r)&=&1 +\frac{ml^3}{(r^{2}+r_{+}^{2})^{n/2}}+%
\frac{\beta ^{2}l^2h(\eta )}{4\pi n(n-1)}. \label{kr2}
\end{eqnarray}
The electromagnetic field equation in the
new coordinate is
\begin{equation}
F_{\phi
r}=\frac{2ql^{n-1}}{(r^{2}+r_{+}^{2})^{(n-1)/2}\sqrt{1-\eta }}.
\label{f33}
\end{equation}

The function $f(r)$ given in Eq. (\ref{F2}) is positive in the whole
spacetime and is zero at $r=0$. One can easily show that the Kretschmann
scalar does not diverge in the range $0\leq r<\infty $. However, the
spacetime has a conic geometry and has a conical singularity at $r=0$,
since
\begin{equation}
\lim_{r\rightarrow 0}\frac{1}{r}\sqrt{\frac{g_{\phi \phi }}{g_{rr}}}\neq 1.
\label{limit}
\end{equation}
That is, as the radius $r$ tends to zero, the limit of the ratio
``circumference/radius'' is not $2\pi$, and therefore the spacetime has a
conical singularity at $r=0$. The canonical singularity can be removed if
one identifies the coordinate $\phi $ with the period
\begin{equation}
\text{Period}_{\phi }=2\pi \left( \lim_{r\rightarrow 0}\frac{1}{r}\sqrt{%
\frac{g_{\phi \phi }}{g_{rr}}}\right) ^{-1}=2\pi (1-4\tau ),
\end{equation}
where $\tau $ is given by
\begin{equation}
\tau  =\frac{1}{4}\left[ 1-\frac{2}{lr_{+}f^{\prime \prime}_{0}}\right].  \label{mu}
\end{equation}
In Eq. (\ref{mu}), $f^{\prime \prime}_0$ is the value
of the second derivative of $f(r)$ at $r=0$, which can be calculated as
\begin{equation}
f^{\prime \prime}_0=2r_{+}^{2}k_{0}^{\prime \prime} \label{fpp}
\end{equation}
where $k_{0}^{\prime \prime }=k^{\prime \prime }(r=0)$.
By the above analysis,
one concludes that near the origin $r=0$ the metric (\ref{Metr1b}) may be written as
\begin{equation}
\frac{r_+^2}{l^2}\left(-dt^2+dX^2\right)+\frac{1}{r_+^4 k^{\prime
\prime}_{0}} \left[dr^2+(l r_+^3 {k^{\prime \prime}}_{0})^2 r^2
d\phi^2\right]\label{defi}.
\end{equation}
This metric describes a spacetime that is locally flat but has a
conical singularity at $r=0$ with a deficit angle $\delta =8\pi
\tau $, which is proportional to the brane tension at $r=0$
\cite{Rand}. Of course, one may ask for the completeness of the
spacetime with $r\geq 0$ (or $\rho \geq r_{+}$). It is easy to see
that the spacetime described by Eq. (\ref{Metr1b}) is both null
and timelike geodesically complete as in the case of
four-dimensional solutions \cite{Lem1,Hor}. In fact, one can show
that every null or timelike geodesic starting from an arbitrary
point can either extend to infinite values of the affine parameter
along the geodesic or end on a singularity at $r=0$ \cite{DB}.

\begin{figure}[ht]
\centering {\includegraphics[width=7cm]{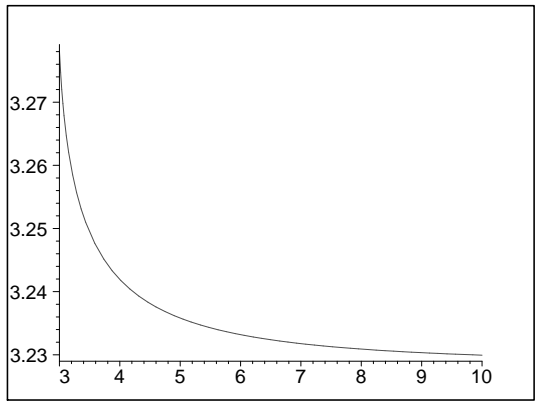} }
\caption{$\delta$ versus $\beta$ for $n=6$, $q =0.2$, $r_{+}=.67$,
$l=1$, $\alpha_2=0.2$, and $\alpha_3=0.1$.} \label{Fig1}
\end{figure}

Now we investigate the effects of different parameters of the
solution on the deficit angle of the spacetime. First, we
investigate the effect of the nonlinearity of the magnetic field
on the deficit angle. In order to do this, we plot $\delta$ versus
the parameter $\beta $. This is shown in Fig. \ref{Fig1}, which shows that as $%
\beta $ decreases (the nonlinearity of the electromagnetic field
becomes more evident) the deficit angle increases. Since for fixed
values of mass, charge, and the cosmological constant the
parameter $\beta $ has a minimum value which can be evaluated by
the fact that $\eta(r=0)<1$, the deficit angle starts from a
maximum value $\delta _{{\rm \max }}$ for $\beta _{\min }$ and
goes to its minimum value as $\beta $ goes to infinity.

Second, we consider the effect of Lovelock terms on the deficit
angle in the following subsection.

\subsection{Static magnetic branes in lower order Lovelock gravity}

As two special cases, we introduce the static magnetic solutions
of Einstein and Gauss-Bonnet gravity in the presence of
a Born-Infeld electromagnetic field. The first special case is when
both of the Lovelock coefficients are zero. Solving Eq.
(\ref{Eqf}) in the case of $\alpha _{2}=\alpha _{3}=0$ and
performing the transformation (\ref{Tr1}), one obtains:
\begin{equation}
f_{\rm Ein}(r) =\frac{r^{2}+r_{+}^{2}}{l^{2}}k(r), \label{Ein}
\end{equation}
where $k(r)$ is given by Eq. (\ref{kr2}) and $r_{+}$ is again the
largest real root of $k(r=0)=0$. Of course, one may also obtain
the above solution by calculating the limit of $f(r)$ of third
order Lovelock gravity given in Eq. (\ref{F2}) as $\alpha_2$ and
$\alpha_3$ go to zero.

As the second special case, we introduce the magnetic solution in
Gauss-Bonnet gravity which may be obtained by solving Eq.
(\ref{Eqf}) in the case of $\alpha_3=0$ and performing the
transformation (\ref{Tr1}). In this case, one obtains
\begin{equation}
f_{{\rm GB}}(r) =\frac{r^{2}+r_{+}^{2}}{2\alpha _{2}}\left( 1-\sqrt{1-\frac{4\alpha_2}{l^2}k(r)}%
\right), \label{GB}
\end{equation}
where again $k(r)$ and $r_+$ are the same as that of Einstein or
Lovelock gravities. Now, we investigate the effects of Lovelock
terms on the properties of the spacetime such as the deficit
angle. According to Eq. (\ref{defi}), the deficit angle depends on
$k_{0}^{\prime \prime }=k^{\prime \prime }(r=0)$ and $r_+$, which
are the same for Einstein, Gauss-Bonnet and third order Lovelock
gravities and are independent of Lovelock coefficients. Thus, one
may expect that Lovelock terms of any order have no effect on
the properties of our spacetime, which is a warped product of an
($n-2$)-dimensional zero-curvature spacetime and a curved
two-dimensional space. In this relation, we will give some
comments in the last section.

\section{Spinning Magnetic Branes\label{Rot}}

Now, we want to endow our spacetime solution (\ref{Met1a}) with a global
rotation. It should be mentioned that the ($t=$const., $r=$const.)-boundary
of our spacetime is curvature-free and therefore this solution is not a
counterpart of the Kerr-type solution in Lovelock gravity.
We first consider the solutions with one rotation parameter. In
order to add angular momentum to the spacetime, we perform the following
rotation boost in the $t$-$\phi $ plane:
\begin{equation}
t\mapsto \Xi t-a\phi \ \ \ \ \ \ \ \ \ \ \phi \mapsto \Xi \phi -\frac{a}{%
l^{2}}t,  \label{tphi}
\end{equation}
where $a$ is the rotation parameter and $\Xi =1+a^{2}/l^{2}$. Substituting
Eq. (\ref{tphi}) into Eq. (\ref{Met1a}) we obtain

\begin{equation}
ds^{2}=-\frac{r^{2}+r^2_+}{l^{2}}\left( \Xi dt-ad\phi \right) ^{2}+\frac{r^2 dr^{2}}{%
(r^2+r^2_+)f(r)}+l^{2}f(r)\left( \frac{a}{l^{2}}dt-\Xi d\phi \right) ^{2}+\frac{r^{2}+r^2_+}{%
l^{2}}dX^{2},   \label{Met2a}
\end{equation}
where $f(r)$ is the same as $f(r)$ given in Eqs. (\ref{F2})-(\ref{kr2}). The
nonvanishing electromagnetic field components become
\begin{equation}
F_{rt}=-\frac{a}{\Xi l^{2}}F_{r\phi }=\frac{2l^{n-3}aq}{r^{n-1}\sqrt{1-\eta }%
}.
\end{equation}
The transformation (\ref{tphi}) generates a new metric, because it is not a
permitted global coordinate transformation. This transformation can be done
locally but not globally \cite{Stach}. Therefore, the metrics (\ref{Met1a}) and (\ref{Met2a}%
) can be locally mapped into each other but not globally, and so they are
distinct. Again, this spacetime has no horizon and curvature singularity.
However, it has a conical singularity at $r=0$.

Second, we study the rotating solutions with more rotation parameters. The
rotation group in $n+1$ dimensions is $SO(n)$, and therefore the number of
independent rotation parameters is $[n/2]$, where $[x]$ is the integer part
of $x$. We now generalize the above solution given in Eq. (\ref{Met1a}) with
$k\leq [n/2]$ rotation parameters. This generalized solution can be
written as
\begin{eqnarray}
ds^{2} &=&-\frac{r^{2}+r_+^2}{l^{2}}\left( \Xi dt-{{\sum_{i=1}^{k}}}a_{i}d\phi
^{i}\right) ^{2}+f(r)\left( \sqrt{\Xi ^{2}-1}dt-\frac{\Xi }{\sqrt{\Xi ^{2}-1}%
}{{\sum_{i=1}^{k}}}a_{i}d\phi ^{i}\right) ^{2}  \nonumber \\
&&+\frac{r^2}{(r^2+r_+^2)f(r)}dr^{2}+\frac{r^{2}+r_+^2}{l^{2}(\Xi ^{2}-1)}{\sum_{i<j}^{k}}%
(a_{i}d\phi _{j}-a_{j}d\phi
_{i})^{2}+\frac{(r^{2}+r_+^2)}{l^{2}}dX^{2}, \label{Met2b}
\end{eqnarray}
where $\Xi =\sqrt{1+\sum_{i}^{k}a_{i}^{2}/l^{2}}$, $dX^{2}$ is the Euclidean
metric on the $(n-k-1)$-dimensional submanifold with volume $V_{n-k-1}$ and $%
f(r)$ is the same as $f(r)$ given in Eq. (\ref{f1}). The nonvanishing
components of the electromagnetic field tensor are
\begin{equation}
F_{rt}=-\frac{(\Xi ^{2}-1)}{\Xi a_{i}}F_{r\phi ^{i}}=\frac{2ql^{n-2}\sqrt{\Xi
^{2}-1}}{r^{n-1}\sqrt{1-\eta }}.
\end{equation}

In the remaining part of this section, we compute the conserved quantities of
the solution. In general, the conserved quantities are divergent when evaluated on the
solutions. A systematic method of dealing with this divergence for
asymptotically AdS solutions of Einstein gravity is through the use of the
counterterms method inspired by the anti-de Sitter conformal field theory
correspondence \cite{Mal}. For asymptotically AdS solutions of
Lovelock gravity with flat boundary $\widehat{R}_{abcd}(\gamma )=0$, the
finite energy-momentum tensor is \cite{DBSH,DM1}
\begin{eqnarray}
T^{ab} &=&\frac{1}{8\pi }\{(K^{ab}-K\gamma ^{ab})+2\alpha'
_{2}(3J^{ab}-J\gamma ^{ab})  \nonumber \\
&&\ +3\alpha'_{3}(5P^{ab}-P\gamma ^{ab})+\frac{n-1}{L}\gamma ^{ab}\ \},
\label{Stres}
\end{eqnarray}
where $L$ is a constant which depends on $l$, $\alpha'_2$, and
$\alpha'_{3}$ that reduces to $l$ as the Lovelock coefficients
vanish. For the special case that $\alpha_{3}=\alpha_{2}^2$, $L$
becomes
\[
L=\frac{15l^{2}\sqrt{\alpha_2 (1-\Gamma )}}{5l^{2}+9\alpha_2
-l^{2}\Gamma ^{2}-4l^{2}\Gamma },
\]
where $\Gamma =(1-3\alpha_2 /l^{2})^{-1/3}$. In Eq. (\ref{Stres}),
$K^{ab}$ is the extrinsic curvature of the boundary, $K$ is its
trace, $\gamma ^{ab}$ is the induced metric of the boundary, and
$J$ and $P$ are traces of $J^{ab}$ and $P^{ab}$ given as
\begin{eqnarray}
J_{ab} &=&\frac{1}{3}%
(2KK_{ac}K_{b}^{c}+K_{cd}K^{cd}K_{ab}-2K_{ac}K^{cd}K_{db}-K^{2}K_{ab}),
\label{Jab} \\
P_{ab} &=&\frac{1}{5}%
\{[K^{4}-6K^{2}K^{cd}K_{cd}+8KK_{cd}K_{e}^{d}K^{ec}-6K_{cd}K^{de}K_{ef}K^{fc}+3(K_{cd}K^{cd})^{2}]K_{ab}
\nonumber \\
&&-(4K^{3}-12KK_{ed}K^{ed}+8K_{de}K_{f}^{e}K^{fd})K_{ac}K_{b}^{c}-24KK_{ac}K^{cd}K_{de}K_{b}^{e}
\nonumber \\
&&+(12K^{2}-12K_{ef}K^{ef})K_{ac}K^{cd}K_{db}+24K_{ac}K^{cd}K_{de}K^{ef}K_{bf}\}.
\label{Pab}
\end{eqnarray}
One may note that, when $\alpha_i$'s go to zero, the finite stress-energy
tensor (\ref{Stres}) reduces to that of asymptotically AdS solutions of
Einstein gravity with a flat boundary.

To compute the conserved charges of the spacetime, we choose a spacelike
surface ${\cal B}$ in $\partial {\cal M}$ with metric $\sigma _{ij}$ and
write the boundary metric in Arnowitt-Deser-Misner form:
\begin{equation}
\gamma _{ab}dx^{a}dx^{a}=-N^{2}dt^{2}+\sigma _{ij}\left( d\varphi
^{i}+V^{i}dt\right) \left( d\varphi ^{j}+V^{j}dt\right) ,
\end{equation}
where the coordinates $\varphi ^{i}$ are the angular variables
parametrizing the hypersurface of constant $r$ around the origin and $N$
and $V^{i}$ are the lapse and shift functions, respectively. When there is a
Killing vector field ${\cal \xi }$ on the boundary, then the quasilocal
conserved quantities associated with the stress tensors of Eq. (\ref{Stres})
can be written as
\begin{equation}
{\cal Q}({\cal \xi )}=\int_{{\cal B}}d^{n-1}\varphi \sqrt{\sigma }T_{ab}n^{a}%
{\cal \xi }^{b},  \label{charge}
\end{equation}
where $\sigma $ is the determinant of the metric $\sigma _{ij}$ and $n^{a}$
is the timelike unit normal vector to the boundary ${\cal B}${\bf .} In the
context of counterterm method, the limit in which the boundary ${\cal B}$
becomes infinite (${\cal B}_{\infty }$) is taken, and the counterterm
prescription ensures that the action and conserved charges are finite. For
our case of horizonless rotating spacetimes, the first Killing vector is $\xi
=\partial /\partial t$, and therefore its associated conserved charge of the
brane is the mass per unit volume $V_{n-k-1}$ calculated as
\begin{equation}
M=\int_{{\cal B}}d^{n-1}x\sqrt{\sigma }T_{ab}n^{a}\xi ^{b}=\frac{(2\pi )^{k}%
}{4}\left[ n(\Xi ^{2}-1)+1\right] m.  \label{Mas}
\end{equation}
The second class of conserved quantities is the angular momentum per unit
volume $V_{n-k-1}$ associated with the rotational Killing vectors $\zeta
_{i}=\partial /\partial \phi ^{i}$, which may be calculated as
\begin{equation}
J_{i}=\int_{{\cal B}}d^{n-1}x\sqrt{\sigma }T_{ab}n^{a}\zeta _{i}^{b}=\frac{%
(2\pi )^{k}}{4}n m\Xi a_{i}.  \label{Ang}
\end{equation}

Next, we calculate the electric charge of the solutions. To determine the
electric field, we should consider the projections of the electromagnetic
field tensor on special hypersurfaces. The normal to such hypersurfaces for
the spacetimes with a longitudinal magnetic field is
\[
u^{0}=\frac{1}{N},\text{ \ }u^{r}=0,\text{ \ }u^{i}=-\frac{N^{i}}{N},
\]
and the electric field is $E^{\mu }=g^{\mu \rho }F_{\rho \nu }u^{\nu }$.
Then the electric charge per unit volume $V_{n-k-2}$ can be found by
calculating the flux of the electromagnetic field at infinity, yielding
\begin{equation}
Q=\frac{(2\pi )^{k}}{2} ql\sqrt{\Xi ^{2}-1}.  \label{elecch}
\end{equation}
Note that the electric charge is proportional to the rotation
parameter and is zero for the case of static solutions. This
quantity and the conserved quantities given by Eqs. (\ref{Mas}) and
(\ref{Ang}) show that the metrics (\ref{Met1a}) and (\ref{Met2b})
cannot be mapped into each other globally.

\section{Closing Remarks}

In this paper, we investigated the effects of the nonlinearity of
the electromagnetic fields on the properties of the spacetime by
finding a new class of magnetic solutions in third order Lovelock
gravity in the presence of a nonlinear Born-Infeld electromagnetic
field. This class of solutions yields an $(n+1)$-dimensional
spacetime with a longitudinal nonlinear magnetic field [the only
nonzero component of the vector potential is $A_{\phi }(r)$]
generated by a static magnetic brane. We found that these
solutions have no curvature singularity and no horizons, but have
conic singularity at $r=0$ with a deficit angle $\delta$ which is
sensitive to the nonlinearity of the electromagnetic field. We
found that as the effects of the nonlinearity of the
electromagnetic fields become larger, the deficit angle increases.
In these static spacetimes, the electric field vanishes, and
therefore the brane has no net electric charge. Next, we endow the
spacetime with rotation. For the spinning brane, when the rotation
parameters are nonzero, the brane has a net electric charge
density which is proportional to the magnitude of the rotation
parameters given by $\sqrt{\Xi ^{2}-1}$. We also applied the
counterterm method in order to calculate the conserved quantities
of the spacetime and found that these conserved quantities do not
depend on the Born-Infeld parameter $\beta $. This can be
understand easily, since at the boundary at infinity, the effects
of the nonlinearity of the electromagnetic fields vanish.

On the effects of Lovelock terms on the properties of our
spacetime, we found that the Lovelock terms have no effects on the
deficit angle of the spacetime. This also happens for the mass,
angular momenta and charge given by Eqs. (\ref{Mas}), (\ref{Ang})
and (\ref{charge}). This feature of our solution is a generic
property of any kind of solutions of any order of Lovelock gravity
for spacetimes which are warped products of an ($n-2$)-dimensional
zero-curvature space (spacetime) and a two-dimensional curved
spacetime (space). For example, all of the thermodynamic quantities
of black holes of Lovelock gravity with a flat horizon, such as the
entropy, the temperature, the free energy, the mass, etc.,  are
independent of the Lovelock coefficients \cite{Maeda}. Of course,
one may note that the thermodynamic quantities of black holes of
Lovelock gravity with a nonflat horizon depend on the Lovelock
coefficients \cite{Maeda}, and therefore one may predict that the
deficit angle of a spacetime with conic geometry which is a warped
product of an ($n-2$)-dimensional nonzero-curvature spacetime and
a two-dimensional curved space should depend on the Lovelock
coefficients. This has been shown for Gauss-Bonnet gravity in the
presence of codimension two branes \cite{Wang}. Finally, we give
some comments on the number of independent solutions in various
order of Lovelock gravity. For static solutions of Lovelock
gravity, the $rr$ components of the field equation can be
integrated in order to get an algebraic equation for $f(r)$. This
algebraic equation is linear for the Einstein equation, quadratic for
Gauss-Bonnet gravity, cubic for third order Lovelock gravity,
quartic for fourth order Lovelock gravity and so on. This suggests
that $f(r)$ has only one branch in Einstein, two branches in
Gauss-Bonnet, three branches in third order Lovelock gravity, and
so on. But the point which should be regarded carefully is the
number of real branches of $f(r)$. Although the number of real
branches is two in Gauss-Bonnet gravity, it is one in third order
Lovelock gravity.

\textbf{Acknowledgements}

This work has been supported by Research Institute for Astrophysics and
Astronomy of Maragha.

\end{document}